\documentclass[showpacs,twocolumn,amsmath,amssymb]{revtex4}
\usepackage{graphicx}
\usepackage{bm}
\begin{document}
\title{Two-fluid behavior of the Kondo lattice in 
the $1/N$ slave boson approach.}
\author{Victor Barzykin} 
\affiliation{Department of Physics and Astronomy, University of Tennessee,
Knoxville, TN  37996-1200}
\begin{abstract}
It has been recently shown by Nakatsuji, 
Pines, and Fisk
[S. Nakatsuji, D. Pines, and Z. Fisk, Phys. Rev. Lett. \textbf{92}, 
016401 (2004)] from the phenomenological analysis of experiments
in Ce$_{1-x}$La$_x$CoIn$_5$ and CeIrIn$_5$ that thermodynamic
and transport properties of Kondo lattices below coherence
temperature can be very successfully described in terms of a
two-fluid model, with Kondo impurity 
and heavy electron Fermi liquid contributions.
We analyze thermodynamic properties of Kondo lattices
using $1/N$ slave boson treatment of the periodic Anderson model
and show that these two contributions indeed arise below the
coherence temperature. We find that the Kondo impurity contribution
to thermodynamics corresponds to thermal excitations into the flat
portion of the energy spectrum. 
\end{abstract}
\vspace{0.15cm}

\maketitle

\section{Introduction}

Anomalous behavior of heavy fermion (HF) metals has attracted enormous theoretical and 
experimental interest over the past 30 years since their discovery\cite{andres}.
Due to the almost localized nature of f-electron bands, electrons in these materials
have strong Coulomb interactions, and exhibit a variety of very unusual properties,
including heavy effective mass of carriers (with $m \sim 100-1000 m_e$),
non-Fermi liquid ground state properties\cite{gorkov},
nontrivial magnetic order (proposed for UPt$_3$, URu$_2$Si$_2$\cite{BGUR,ramcolch,colch}), and unconventional superconductivity,
observed in most, if not all of HF superconductors (see, for example, Refs \cite{gorkov, SU} for a review).
Recent theoretical and experimental interest in these materials has been focused on the quantum
phase transitions. 
Quantum critical anomalous power law behavior with temperature of thermodynamic, 
magnetic, and transport properties\cite{coleman}  is observed in the vicinity of a quantum phase transition
in HF compounds such as CeCu$_2$Si$_2$\cite{steglich}, CeCu$_{5.9}$Au$_{0.1}$\cite{schroder}, YbRh$_2$Si$_2$\cite{steglich},
or CeCoIn$_5$ and CeIrIn$_5$\cite{cedomir,movshovich,NPF}. 
The surprising experimental discovery is that the quantum criticality in heavy fermion metals is
also unconventional, i.e., it does not follow the usual Hertz-Millis\cite{herts, millis} scaling analysis of
a magnetic quantum critical point. Instead, the fluctuations remain local,
as it was demonstrated in the inelastic neutron scattering experiments\cite{schroder}, 
which strongly suggests a Kondo-type local quantum criticality\cite{coleman,si}. 

The most interesting question remains the formation and the origin of
heavy mass in these materials. Nakatsuji, Pines, and Fisk\cite{NPF}(NPF)
have recently shown that our understanding of Kondo lattices is incomplete.
NPF propose that formation of heavy mass in heavy fermion
materials (in particular, in Ce$_{1-x}$La$_x$CoIn$_5$ and CeIrIn$_5$), which happens at a crossover temperature $T^*$, 
can be described by a phenomenological theory similar in spirit to Landau theory of superfluidity\cite{LL9}.
The NPF two-fluid phenomenological model\cite{NPF,NPFcurro} states that thermodynamic, magnetic, and transport properties of Kondo lattices
can be described as a sum of two independent contributions, one involving the HF liquid, the other 
a lattice of weakly interacting Kondo impurity centers. The relative fraction $f$ of the coherent HF state plays the role of the order
parameter, which develops below the coherence temperature, $T^*$. 

We demonstrate below that the same type of phenomenology directly follows from the lattice
analogue of the slave boson broken symmetry ansatz of Read and Newns\cite{readn} and Coleman\cite{col}. 
The order parameter,  the   slave boson average, gives rise to the coherent heavy fermion liquid component, 
while the contribution from almost flat f-electron bands accounts for the  Kondo impurity component.
 Although no actual symmetry breaking occurs in this case,
an analog of Landau two-fluid model in superconductors arises in the slave boson language. The
 heavy Fermion liquid contribution corresponds to the bose-condensate of slave bosons, while the Kondo impurity
contribution is a complete analogue of the normal part\cite{LL9}.

While the slave boson language provides a very attractive physical explanation of the NPF phenomenology\cite{NPF},
in does not completely agree with it. 
 We provide a detailed comparison of the slave boson results  with the NPF phenomenology in the summary.          

\section{The slave boson method}

Theoretical approach to understanding 
the physics of HF usually starts with the Kondo lattice, or more general periodic Anderson\cite{Anderson} model Hamiltonian,
\begin{equation}
H = H_0 + H_V,
\label{theH}
\end{equation}
where
\begin{widetext}
\begin{equation}
H_0 = \sum_{\bm{k} \sigma} \epsilon_k c^{\dagger}_{\bm{k} \sigma} c_{\bm{k} \sigma} + 
\sum_{i \sigma} E_0 f^{\dagger}_{i \sigma} f_{i \sigma} +
\sum_i U f^{\dagger}_{i \uparrow} f^{\dagger}_{i \downarrow} f_{i \downarrow} f_{i \uparrow}. 
\label{fterm}
\end{equation}
\end{widetext}
Here the creation and annihilation operators for the f-electrons, $f^{\dagger}_{i \sigma}$ and $f_{i \sigma}$,
carry the site index $i$, and there is a Coulomb interaction at each site for the f-electrons. The operators
$c^{\dagger}_{\bm{k} \sigma}$ and $c_{\bm{k} \sigma}$ correspond to delocalized Bloch states.
The hybridization term $H_V$ accounts for the $c-f$ hybridization between the  f-electrons and
the Bloch states:
\begin{equation}
 H_V = \sum_{i, \bm{k}, \sigma} 
 V_{\bm{k}} e^{i \bm{k} \cdot \bm{R}_i} f^{\dagger}_{i \sigma} c_{\bm{k} \sigma}  +
 V_{\bm{k}}^* e^{-i \bm{k} \cdot \bm{R}_i} c^{\dagger}_{\bm{k} \sigma} f_{i \sigma}
\label{hyb}
\end{equation}
Both Anderson and Kondo models have been solved exactly for a single f-electron site (for a review, see Ref.\cite{hewson,TW}).
Even though the Bethe ansatz solution is by no means easy,
most of our basic theoretical knowledge about Kondo systems is inferred from it. Unfortunately,
exact solution cannot be obtained for a periodic lattice,
where correlations involve several conduction and f-electron spins. Nevertheless,
it can be treated quite successfully using various approximate methods, such as the non-crossing approximation, 
the $1/N$ approach\cite{hewson}, or dynamical mean field theory\cite{kotliarrmp}.      

The large-$N$ slave boson approach was
developed in the mid-80-s by a number of authors\cite{read,col,readn,barnes} for single impurity Kondo and Anderson models, 
and was later applied to lattices \cite{cole,readne,mil:lee,levin}. 
The result of this approach for the
single impurity models has been shown to approximate well exact results at both low and high temperatures\cite{coland}.
Following these references, we can rewrite the periodic Anderson model Eq.(\ref{theH}) in the slave boson
language.
Since the on-site Coulomb repulsion $U$ is very large for f-electron materials, it is usually taken as
infinite. Then the creation and annihilation operators for $f$-electrons need to be applied with
special projection operators, which project out doubly occupied $f$-electron states\cite{read}. A convenient way
to deal with this Hamiltonian \cite{col,readn,barnes} is to introduce
a new slave boson field $b_i^+$, which creates a hole on site $i$, and to rewrite the Anderson Hamiltonian in a
way which allows the $1/N$ expansion in the number of orbitals. As a result, the infinite-$U$ Anderson model 
takes the following form:
\begin{eqnarray}
H & = & \sum_{\bm{k} m} \left[ \epsilon_k c^{\dagger}_{\bm{k} m} c_{\bm{k} m} + E_0 f^{\dagger}_{\bm{k} m} f_{\bm{k} m} \right] + \nonumber \\ 
& + & \sum_{\bm{k}, \bm{q}, m} \left( c^{\dagger}_{\bm{k} m} f_{\bm{q} m} b^{\dagger}_{\bm{q} - \bm{k}} + h.c. \right).
\label{slb}
\end{eqnarray}   
Here $m$ is a spin index, which runs from $1$ to $N$ (i.e., we assume $N$ degenerate f-levels \textbf{and} $N$ conduction bands),
and the sums for $\bm{k}$ and $\bm{q}$ run over the Brillouin zone.
The absence of doubly occupied states is ensured by a constraint\cite{read} for every $f$-site $i$:
\begin{equation}
 Q_i = \sum_{m} f^{\dagger}_{im} f_{im} + b^{\dagger}_i b_i = 1.
\label{constr}
\end{equation}
The large-$N$ slave boson formulation and the $1/N$ expansion around it have many well-known difficulties.
Nevertheless, the results of this model  capture the low temperature heavy Fermi liquid properties of Kondo lattices rather well,
and agree with many experiments\cite{hewson}. The mean field approach starts with
a replacement of the slave boson operators with their averages \cite{col, read, mil:lee},
\begin{equation}
\langle b_i \rangle = a
\label{brs}
\end{equation} 
After taking into account the constraint Eq.(\ref{constr}), the resulting Hamiltonian then can be written as:
\begin{eqnarray}
H &=& \sum_{\bm{k}, m} \left[\epsilon_{\bm{k}} c^{\dagger}_{\bm{k} m} c_{\bm{k} m}  
+ \epsilon_f f^{\dagger}_{\bm{k} m} f_{\bm{k} m} \right. \nonumber \\
&+&
\left. (Va)(f^{\dagger}_{\bm{k} m} c_{\bm{k}m} + h.c.) + (\epsilon_f - E_0)(a^2 -1)\right],
\label{MFH}
\end{eqnarray}
where $\epsilon_f$ is the renormalized $f$-energy level. The parameters $\epsilon_f$ and $a$ are fixed
by the mean field equations, which are obtained by minimizing the free energy.

A serious well-known limitation of the $1/N$ slave boson approach arises from the necessity to satisfy the 
constraint Eq.(\ref{constr}). 
According to quantum mechanics, the phase of the boson field is conjugate to the number operator
for it. Thus, a state in which the boson has a definite phase, such as one introduced by the usual slave boson mean field ansatz 
Eq.(\ref{brs}), 
results in infinite uncertainty for the boson number operator, and necessarily violates the local constraint
Eq.(\ref{constr}). This introduces unphysical divergences in correlation functions, which don't have direct physical meaning.
These divergences must cancel for all physical properties.   According to Read\cite{read1}, divergent goldstone mode
phase fluctuations act to restore the true symmetry of the system, and $\langle b_i \rangle = 0$. These phase fluctuations
can be integrated out in the functional integral formalism, while the ansatz Eq.(\ref{brs}) will hold for the boson amplitude.
In this sense, this phase transition is similar to the Kosterlitz-Thauless phenomenon. In agreement with
x-ray orthogonality catastrophe\cite{ND}, the local spin correlations
in the single impurity model are critical, i.e., have a power law decay at large times \cite{cole}.
Fluctuations in the constraint lead to large deviations from mean field at temperatures $T \sim T_K$.
Nevertheless, if the limit $N \rightarrow \infty$ is taken at finite filling $q_0$,
the mean field solution for a single impurity Anderson model approaches
the Bethe ansatz solution at both high and low temperature end \cite{coland}, 
with most deviations occurring around the mean field transition temperature $T^* \sim T_K/\ln{N}$. The convergence is
excellent at low temperatures, although it remains rather poor at $T > T^*$ \cite{coland}. 
This well-known problem with the slave boson method has prevented
the development of the two-fluid microscopic model in the past. In particular, 
the Kondo gas thermodynamic contribution from almost flat band, which appears already in the tree approximation, has always
been dismissed as small, as an artifact of the above mentioned problem with the constraint, Eq.(\ref{constr}). 

The mean field Hamiltonian Eq.(\ref{MFH}) can be diagonalized by the canonical transformation, which
results in a well known\cite{readn} spectrum with a gap,
\begin{eqnarray}
\epsilon_{1,2}(\bm{k}) &=& \frac{1}{2}\, (\epsilon_{\bm{k}} + \epsilon_f \pm E_{\bm{k}}) \label{sp1} \\
E_{\bm{k}} &=& \sqrt{(\epsilon_{\bm{k}} - \epsilon_f)^2 + 4 (Va)^2} \label{sp2} 
\end{eqnarray}
The new energy spectrum gives rise to the mass enhancement given by
\begin{equation}
\frac{m^*}{m}\, = \frac{V^2 a^2}{\epsilon_f^2}\, + 1,
\label{mass}
\end{equation}
which becomes rather dramatic in the Kondo regime.

It is convenient to introduce re-defined bare-fermion Green's functions, 
\begin{eqnarray}
G_{cc}^m(\bm{k}, \tau) &\equiv& - \langle T_{\tau} c_{\bm{k} m} (\tau) c^{\dagger}_{\bm{k} m} (0) \rangle_{MF}, \\
G_{fc}^m(\bm{k}, \tau) &\equiv& - \langle T_{\tau} f_{\bm{k} m} (\tau) c^{\dagger}_{\bm{k} m} (0) \rangle_{MF}, \\
G_{ff}^m(\bm{k}, \tau) &\equiv& - \langle T_{\tau} f_{\bm{k} m} (\tau) f^{\dagger}_{\bm{k} m} (0) \rangle_{MF},
\end{eqnarray}
where the thermal average is taken in a system described by Eq.(\ref{MFH}) and $\tau$ is imaginary time.
Transforming to Matsubara frequencies, it is easy to find:
\begin{eqnarray}
\label{verygreen}
G_{cc}^m(\bm{k}, \omega_n) &=& \frac{i \omega_n - \epsilon_f}{(i \omega_n - \epsilon_f)(i \omega_n - \epsilon_{\bm{k}}) - V^2 a^2}\, \\
G_{ff}^m(\bm{k}, \omega_n) &=& \frac{i \omega_n - \epsilon_{\bm{k}}}{(i \omega_n - \epsilon_f)(i \omega_n - \epsilon_{\bm{k}}) - V^2 a^2}\, \\
G_{fc}^m(\bm{k}, \omega_n) &=& \frac{Va}{(i \omega_n - \epsilon_f)(i \omega_n - \epsilon_{\bm{k}}) - V^2 a^2}\,
\end{eqnarray}
The mean field parameters are determined by minimizing the free energy for the Hamiltonian in Eq.(\ref{MFH}), which
diagrammatically leads to vanishing of the sum of the ``tadpole'' graphs\cite{mil:lee},
\begin{eqnarray}
\label{first}
N (\epsilon_f - E_0) a + V T \sum_{\bm{k},n,m} G_{fc}^m(\bm{k}, \omega_n) &=& 0 \\
N(a^2 - q_0) + T \sum_{\bm{k},n,m} G_{ff}^m(\bm{k}, \omega_n) &=& 0,
\label{second}
\end{eqnarray}
where $q_0$ is the filling factor, $q_0 = 1/N$. 
For free f-electrons above $T^*$, Eq.(\ref{second}) gives:
\begin{equation} n_f(\epsilon_f) = q_0 = \frac{1}{N}\,, \end{equation}
so that the entropy for the f-levels
\begin{equation} S =  - N (q_0 \ln{q_0} + (1-q_0) \ln{(1-q_0)} ) \end{equation}
Setting the filling $q_0 = 1/N$  gives
$S \simeq \ln{N} + 1$.
Exact entropy at high temperatures is not hard to find - one has N half-filled bands and $\ln{N}$ coming from each $N$-fold degenerate
f-level. Thus,
$S_{exact} = \ln{N},$
and we make a mistake of the order of $1/\ln{N}$ in the entropy when we do mean field.
In Coleman approach\cite{cole} the $1/N$ expansion takes place at a given $q_0$, and
$q_0$ is set to $1/N$ in the end. Formally this gives a correction 
\begin{equation}
S_{exact} - S \simeq - 0.5 \ln{(2 \pi N q_0 [1-q_0])},
\end{equation} 
which is $O(\ln{N}/N)$, which makes such $1/N$ expansion somewhat better suited for calculating finite temperature 
properties\cite{coland}. However, the two-component physics is not evident in the Coleman approach. The reason
is that the Fermi liquid Sommerfeld expansion, which appears as an expansion in $1/\ln{N}$ in the Read and
Newns \cite{readn} approach, becomes divergent at $T \sim T^*$ in the Coleman approach. Hence, the Fermi
liquid contribution cannot be easily identified.
In what follows,
we use the approach of Read and Newns\cite{readn} to calculate two-component thermodynamics at the mean field level.

\section{The mean field equations.}

The mean field equations Eqs(\ref{first}),(\ref{second}), together with the conservation of the 
total number of particles,
\begin{equation}
 N_{tot} = T \sum_{\bm{k},n,m}\left[ G_{ff}^m(\bm{k}, \omega_n) + G_{cc}^m(\bm{k}, \omega_n) \right] = const,
\label{third}
\end{equation}
define the order parameter for the second order phase transition in the slave boson method. 

Transforming the integration variable in the mean field equations  Eqs(\ref{first}),(\ref{second})to 
the new energy spectrum Eq.(\ref{sp1}), $\epsilon = \epsilon_{1,2}(\bm{k})$, we get\cite{readn}:

\begin{equation}
\epsilon_f - E_0 = \rho_0 V^2  \left[ \int_{\epsilon_a}^{\epsilon_b} + \int_{\epsilon_c}^{\epsilon_d} \right]
 \frac{n_f(\epsilon)}{\epsilon_f - \epsilon}\, d \epsilon
\label{first1}
\end{equation}
\begin{equation}
\frac{1}{N a^2}\, = 1 +  
\rho_0 V^2 \left[ \int_{\epsilon_a}^{\epsilon_b} + \int_{\epsilon_c}^{\epsilon_d} \right]
\frac{n_f(\epsilon)}{(\epsilon_f - \epsilon)^2}\, d \epsilon.
\label{second1}
\end{equation}
The limits for the integrals are easily calculated from the energy spectrum, Eqs(\ref{sp1}),(\ref{sp2}):
\begin{eqnarray}
\epsilon_a & = & - D - \frac{(Va)^2}{D + \epsilon_f}\, \nonumber \\
\epsilon_b & = & \epsilon_f - \frac{(Va)^2}{D - \epsilon_f}\, \nonumber \\
\epsilon_c & = & \epsilon_f + \frac{(Va)^2}{D + \epsilon_f}\, \nonumber \\
\epsilon_d & = & D + \frac{(Va)^2}{D - \epsilon_f}\, 
\label{coffs}
\end{eqnarray}

The effective Kondo temperature $\epsilon_f$ can be calculated from Eq.(\ref{first1}). This integral is the
Cauchy limit integral around the Kondo gap, which is not divergent. 
In the Kondo limit, $E_0 < 0$, $|E_0| \gg V$,
the zero-temperature value of $\epsilon_f$, $\epsilon_{f0}$, is easily calculated from Eq.(\ref{first1}):
\begin{equation}
\epsilon_{f0} = D e^{-\frac{|E_0|}{\rho_0 V^2}\,}
\end{equation}
Now, for $T \neq 0$, but $T \ll \epsilon_f$, one can use the Sommerfeld expansion to get:
\begin{equation}
|E_0| = \rho_0 V^2 \ln{\left[ \frac{D}{\epsilon_f}\, \right]} + \frac{\pi^2}{6}\, \rho_0 V^2 \frac{1}{(\beta \epsilon_f)^2}\,,
\end{equation}
or, equivalently,
\begin{equation}
\epsilon_f(T) \simeq \epsilon_{f0} \left(1 + \frac{\pi^2}{6}\, (\beta \epsilon_{f0})^{-2} \right).
\label{ef}
\end{equation}

The second mean field equation, Eq.(\ref{second1}), determines the order parameter $a$, or the Kondo gap.
The integral diverges strongly as $(\epsilon-\epsilon_f)^{-2}$  at $\epsilon \rightarrow  \epsilon_f$. 
To deal with this divergence, we can integrate by parts: 
\begin{widetext}
\begin{equation}
\frac{1}{N a^2}\, = 1 - \rho_0 \frac{V^2}{D}\, + \frac{\rho_0 D}{a^2}\, 
\left[n_f\left(\epsilon_f - \frac{V^2 a^2}{D}\,\right) +  n_f\left(\epsilon_f + \frac{V^2a^2}{D}\,\right)\right] +
\rho_0 V^2 \left[\int_{\epsilon_a}^{\epsilon_b} + \int_{\epsilon_c}^{\epsilon_d} \right] 
\frac{n_f'(\epsilon) d \epsilon}{\epsilon - \epsilon_f}. 
\label{gap0}
\end{equation}
\end{widetext}
The latter integral is now convergent in Kondo gap region, so it can be taken using Sommerfeld expansion,
\begin{equation}
\rho_0 V^2 \left[\int_{\epsilon_a}^{\epsilon_b} + \int_{\epsilon_c}^{\epsilon_d} \right] 
\frac{n_f'(\epsilon) d \epsilon}{\epsilon - \epsilon_f} \simeq \frac{\rho_0 V^2}{\epsilon_f}\, 
\left[ 1 + \frac{\pi^2 T^2}{3 \epsilon_f^2}\, \right] 
\end{equation}
To solve for $T^*$, we take the limit $a^2 \rightarrow 0$ in Eq.(\ref{gap0}):
\begin{equation}
q_0 =  2 \rho_0 D n_f(\epsilon_{f}),
\end{equation}
which gives
\begin{equation}
T^* \simeq \frac{\epsilon_{f}(T^*)}{\ln{\left[2 \rho_0 D N\right]}}\,
\label{tc}
\end{equation}
The zero-temperature value of the order parameter $a$, $a(T=0) \equiv a_0$, can also be easily found from Eq.(\ref{gap0}):
\begin{equation}
a_0^2 = \frac{1}{N}\, \frac{\epsilon_{f0}}{\epsilon_{f0} + \rho_0 V^2}\, 
\end{equation}
The transcendental equation Eq.(\ref{gap0}), which gives the temperature dependence of the order parameter $a$,
is considerably simplified if we only calculate the lowest terms in $1/\ln{N}$. Indeed, since $T < T^* \sim \epsilon_f/\ln{N}$,
the asymptotic Sommerfeld expansion becomes an expansion in $1/\ln{N}$. To keep the lowest order in $1/\ln{N}$ in the
Sommerfeld expansion, we take
\begin{equation}
\epsilon_f (T < T^*) \simeq \epsilon_{f0}
\end{equation}
in Eq.(\ref{tc}).
One can simplify the gap equation Eq.(\ref{gap0}) in the Kondo regime, where 
\begin{equation}
\Delta_k = \frac{\rho_0 V^2 a^2}{D}\, \ll \epsilon_f.
\end{equation}
Assuming a square band $2 \rho_0 D = 1$, following Newns and Read\cite{readn}, keeping the leading terms in $O(1/\ln{N})$,
\begin{equation}
\frac{a^2(T)}{a_0^2}\, = 1 - N n_f(\epsilon_{f})\,
\label{haha}
\end{equation}
The third equation, Eq.(\ref{third}), determines the shift of the chemical potential in the slave boson phase transition.
In the Kondo limit the number of electrons per $f$-level approximately stays at $1$, and no significant shift of the chemical potential
occurs.

\section{Thermodynamics of the Kondo lattice - the two-component picture.}

Thermodynamic properties can easily be calculated from the mean field free energy for the Hamiltonian Eq.(\ref{MFH}).
From Luttinger functional, we can write for the Grand free energy density without the small free conduction electron contribution:
\begin{equation}
\Omega = - P = \sum_k \Omega_f + \delta \Omega + \sum_k N (a^2 - q_0)(\epsilon_f - E_0),
\label{therm}
\end{equation}
where $\Omega_f$ is the grand free energy density for free non-hybridized f-centers, $\delta \Omega$
is the effect of hybridization (mixed contribution).
Since $\Omega_f$ corresponds to a free fermion at renormalized energy $\epsilon_f$ with respect to chemical potential,
\begin{equation}
\Omega_f = - N T \ln\left( 1 + e^{- \beta \epsilon_f} \right). 
\end{equation}
The third term in Eq.(\ref{therm}) comes from the energy shift due to a slave boson average. 
The interaction energy $\delta \Omega$ can be expressed through the mixed Green's function,
\begin{equation}
\delta \Omega = 2 V \int_0^a da T \sum_{\bm{k}, n, m} G_M^m(i \omega_n, \bm{k}).
\end{equation}
 We note that the RHS of this equation corresponds to the Green's function expression in Eq.(\ref{second}). 
We can use Eq.(\ref{second}) to replace the expression under the integral. Since from Eq.(\ref{second})
\begin{equation}
V T \sum_{\bm{k}, n, m} G_M^m(i \omega_n, \bm{k}) = - c_f N (\epsilon_f - E_0) a ,
\end{equation}
\begin{equation}
\delta \Omega = - 2 c_f N \int_0^a (\epsilon_f - E_0) a da  = - N c_f a^2 (\epsilon_f - E_0),
\end{equation}
where $c_f$ is the density of f-levels. As a result, we easily obtain:
\begin{equation}
\Omega = -  N c_f T \ln\left( 1 + e^{- \beta \epsilon_f} \right) -  N c_f q_0 \epsilon_f 
\end{equation}
Houghton, Read, and Won\cite{HRW} have found that a similar expression for the Grand free energy
for a single-impurity model holds to two loops.
Thermodynamics is obtained, as usual, by differentiating the Grand Free energy with respect to temperature:
\begin{widetext}
\begin{equation}
S = - \frac{\partial \Omega}{\partial T}\, = N c_f \ln{\left(1 + e^{- \beta \epsilon_f} \right)} + c_f \frac{d \epsilon_f}{d T}\, (1 - N n_f(\epsilon_f)) +
\beta \epsilon_f N c_f n_f(\epsilon_f) 
\end{equation}
\end{widetext}
Comparing this with expression Eq.(\ref{haha}) for $a^2$, and keeping the terms of lowest order in $O(1/\ln{N})$ (the first term
is then small compared with the last term), we can write:
\begin{equation}
S = \beta \epsilon_f c_f \frac{a_0^2 - a^2}{a_0^2}\, +
\frac{\pi^2}{3}\, N \rho_0 \frac{V^2 a^2}{\epsilon_f^2}\, T, 
\end{equation}
or
\begin{equation}
S \simeq \gamma_{HFL} T f(T) + \beta \epsilon_f c_f (1 - f(T)),
\end{equation}
where 
\begin{equation}
f(T) = \frac{a(T)^2}{a_0^2}\,,
\label{fract}
\end{equation}
and $\gamma_{HFL} = \frac{2 \pi^2}{3}\, N \rho_0 \frac{m^*}{m}\,$.
The heat capacity can be found by differentiating the entropy $S$:
\begin{widetext}
\begin{equation}
C = T \frac{d S}{d T}\, = c_f \frac{\epsilon_f^2}{T^2}\, \frac{a_0^2 - a^2}{a_0^2}\, +
\frac{\pi^2}{3}\, N \rho_0 \frac{V^2 a^2}{\epsilon_f^2}\, T 
\end{equation}
\end{widetext}
where we have neglected all terms smaller in $T/\epsilon_f \sim 1/\ln{N}$ in the first part,
although all of them can be written out in the same fashion:
\begin{equation}
C = T \frac{d S}{d T}\, = C_1 c_f \frac{a_0^2 - a^2}{a_0^2}\, + C_{FL} \frac{a^2}{a_0^2}\,.
\end{equation}
Here $C_{FL}$ is the usual heavy Fermi liquid contribution, 
\begin{equation}
C_{FL} =   \gamma_{HFL} T,
\end{equation}
while
\begin{equation}
C_1 \simeq  \frac{\epsilon_f^2}{T^2}\,,
\end{equation}
a factorized contribution. We note that unlike Ref.\cite{NPF}, the dominant contribution in $1/\ln{N}$
comes from differentiating the order parameter $a^2(T)$, 
\begin{equation}
\frac{d (a^2)}{dT}\, \simeq - \frac{\epsilon_f}{T^2}\, (a^2 - a_0^2).
\end{equation} 

In magnetic field we can write, neglecting free conduction electron contribution:
\begin{equation}
\Omega = - T c_f \sum_m \ln{\left(1 + e^{- \beta (\epsilon_f + h m)} \right)}  - c_f \epsilon_f
\end{equation}
Now, $N$ in MF equations is replaced by sum over different $\epsilon_f(h,T) + h m$.
Taking a partial derivative with respect to $h = g \mu_B B$, we find the magnetic moment:
\begin{equation}
\frac{M}{g \mu_B}\, = - c_f \sum_m \left(\frac{\partial \epsilon_f}{\partial h} + m \right) n_f(\epsilon_f + hm) 
+ c_f \frac{\partial \epsilon_f}{\partial h} 
\end{equation}
For small h,
\begin{equation}
\frac{M}{g \mu_B}\, = c_f \frac{\partial \epsilon_f}{\partial h} (1 - N n_f(\epsilon_f)) - h c_f \sum_m m^2 n_f'(\epsilon_f) 
\end{equation}
This obviously separates into two contributions. After some simple algebra, we find that the bulk spin susceptibility
also has two separate contributions:
\begin{equation}
\chi \simeq \chi_{HFL} \frac{a^2}{a_0^2}\, + \frac{(g \mu_B)^2 j (j+1) c_f}{3T}\, \left(1 - \frac{a^2}{a_0^2}\, \right),
\end{equation}
where 
\begin{equation}
\chi_{HFL} = \frac{N (g \mu_B)^2 j (j+1)}{3}\, \rho_0 \frac{m^*}{m}\,,
\end{equation}
and $m^*/m$ is the mass enhancement at $T=0$, $j=\frac{N-1}{2}\,$. The second term on the rhs corresponds to the
typical Curie impurity contribution. A correct logarithmic correction to it appears in the next order in $1/N$\cite{bruls} 
 
\begin{figure}
\includegraphics[width=2.5in]{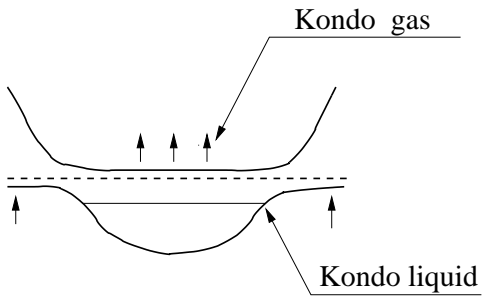}
\caption{The heavy liquid component comes from the renormalized band contribution, while the gas component appears due
to thermal population of almost localized part of the spectrum.} 
\label{2comp}
\end{figure}

The origin of the two components in thermodynamics
is shown schematically on Fig.\ref{2comp}. Indeed, 
the new energy spectrum for free fermions gives a singularity in density of states,
\begin{equation}
\rho(\epsilon) = \rho_0 \left[1 + \frac{(Va)^2}{(\epsilon - \epsilon_f)^2}\, \right].
\end{equation}
This singularity contributes to thermodynamic properties. Since the mean field equations are derivatives of the free energy, 
the singularities that 
are encountered are of the same type as those for 
$\epsilon_f$ and $a^2$. Therefore, the final results can be expressed in terms of $a^2(T)$.

\section{Summary}

\begin{figure}
\includegraphics[width=2.5in]{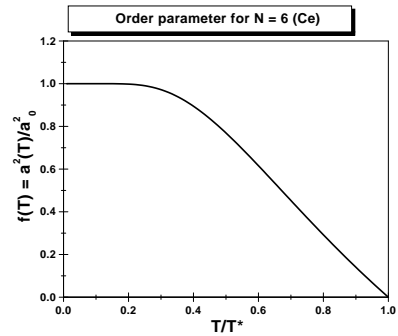}
\caption{The temperature dependence $f(T) = \frac{a^2(T)}{a_0^2}\,$ for Ce (N=6)} 
\label{order}
\end{figure}

Comparing the mean field results with experimental phenomenology of Ref. \cite{NPF}, we find
many similarities. For example, thermodynamics for the Kondo lattice in the slave boson approach clearly factorizes
into the Kondo impurity and heavy electron liquid contributions. Similar to situation in superconductivity
this factorization is a direct result of the appearance of anomalous mixed averages of the type $\langle c^{\dagger} f \rangle$,
which produces the energy gap, or the mean field order parameter $a^2$, which corresponds to
the fraction of the Kondo liquid, Eq.(\ref{fract}).
The linear behavior of $f(T)$ for $T < T^*$, which is different from that
given by Eq.(\ref{haha}), was found by fitting experiments in the 1-1-5 family in Ref.\cite{NPF}. 
In case of $Ce$ ($N=6$) we find that numerically Eq.(\ref{haha}) gives
an order parameter which is almost linear (See Fig.\ref{order})
The spin susceptibility  $\chi(T)$ follows the experimental behavior, and the 
behavior of thermodynamic functions factorizes into two components as well. Clearly
this is a very rough approximation, and many effects observed experimentally in Ref.\cite{NPF} 
are not captured. For example,
at zero temperature we find that $f(T=0) = 1$, and the Kondo impurity contribution is absent, unlike the limit $f(T=0) \simeq 0.9$ found
in Ref. \cite{NPFcurro}. This discrepancy could, in principle, be explained by disorder, which would produce
a finite population of the DOS peak at zero temperature. We also note that the above model is considerably 
oversimplified. For example, the $\bm{k}$-dependence of $V$ was neglected.
In a realistic case $V(\bm{k})$ may have zeroes on the Fermi surface, which will lead to 
a $k$-dependent energy gap with a nodal structure, similar to that observed in unconventional superconductors.

We also find that thermodynamics ($C$, $S$, etc.) does not completely follow the NPF phenomenology.
While the separation in two components is always present, we found that it is necessary to differentiate
$f(T)$ to get the correct answer for some quantities. It is also, perhaps, necessary to get to the next approximation,
in order to find the log(T) dependence of effective mass, observed in Ref.\cite{NPF}. Calculation
of transport properties requires going beyond the mean field approximation \cite{mil:lee}, which, in principle,
could also explain $f(T=0) \neq 1$ without introducing disorder. Nevertheless, this approach can be regarded as
the first approximation to understanding experimental results.

I would like to thank L.P. Gor'kov for many discussions and insights.
This work was supported by TAML at the University of Tennessee.

\end{document}